\documentstyle{cupconf}
\input epsf

\ifoldfss
\else
  \ifnfssone
    \newmathalphabet{\mathit}
      \addtoversion{normal}{\mathit}{cmr}{m}{it}
      \addtoversion{bold}{\mathit}{cmr}{bx}{it}
    \newmathalphabet{\mathcal}
      \addtoversion{normal}{\mathcal}{cmsy}{m}{n}
    \else
    \ifnfsstwo
    \fi
  \fi
\fi

\def\hexnumber#1{\ifcase#1 0\or1\or2\or3\or4\or5\or6\or7\or8\or9\or
 A\or B\or C\or D\or E\or F\fi }

\makeatletter
\ifx\CUP@mtlplain@loaded\undefined
\else
\fi
\makeatother
 \makeatletter
 \ifx\CUP@mtlplain@loaded\undefined
   \font\tenbmi=cmmib10 at 10pt
   \font\sevenbmi=cmmib10 at 7pt
   \font\fivebmi=cmmib10 at 5pt

   \newfam\bmifam
   \textfont\bmifam=\tenbmi
   \scriptfont\bmifam=\sevenbmi
   \scriptscriptfont\bmifam=\fivebmi
   
 \fi
 \makeatother
\ifnfsstwo

\fi
\ifnfssone

\fi
\ifoldfss

\fi

\mathchardef\varLambda="0103

\makeatletter
\ifx\CUP@mtlplain@loaded\undefined
\else
\fi
\makeatother
\makeatletter
\ifx\CUP@mtlplain@loaded\undefined
  \font\tenbms=cmbsy10
  \font\sevenbms=cmbsy10 at 7pt
  \font\fivebms=cmbsy10 at 5pt
  \newfam\bmsfam
  \textfont\bmsfam=\tenbms
  \scriptfont\bmsfam=\sevenbms
  \scriptscriptfont\bmsfam=\fivebms

  \edef\bsy@{\hexnumber\bmsfam}
  \mathchardef\bnabla="0\bsy@72
\fi
\makeatother
\title[Atomic Gas in Spiral Galaxies]{Properties of Atomic Gas 
in Spiral Galaxies}

\author[R. Braun]%
{R\ls O\ls B\ls E\ls R\ls T\ns B\ls R\ls A\ls U\ls N$^1$}

\affiliation{$^1$Netherlands Foundation for Research in Astronomy,
Postbus 2, 7990AA Dwingeloo, The Netherlands}

\begin{document}
\ifnfssone
\else
  \ifnfsstwo
  \else
    \ifoldfss
      \let\mathcal\cal
      \let\mathrm\rm
      \let\mathsf\sf
    \fi
  \fi
\fi

\maketitle

\begin{abstract}
Physical properties of the atomic gas in spiral galaxies are briefly
considered. Although both Warm (WNM, 10$^4$~K) and Cool (CNM, $\sim$~100
K) atomic phases coexist in many environments, the dominant mass
contribution within a galaxy's star-forming disk (R$_{25}$) is that of the
CNM. Mass fractions of 60 to 90\% are found within R$_{25}$. The CNM is
concentrated within moderately opaque filaments with a face-on surface
covering factor of about 15\%. The kinetic temperature of the CNM
increases systematically with galactocentric radius, from some 50 to
200~K, as expected for a radially declining thermal pressure in the
galaxy mid-plane. Galaxies of different Hubble type form a nested
distribution in T$_K$(R), apparently due to the mean differences in
pressure which result from the different stellar and gas surface
densities. The opaque CNM disappears abruptly near R$_{25}$, where the
low thermal pressure can no longer support the condensed atomic
phase. The CNM is apparently a prerequisite for star
formation. Although difficult to prove, all indications are that at
least the outer disk and possibly the inter-arm atomic gas are in the
form of WNM, which accounts for about 50\% of the global total. Median
line profiles of the CNM display an extremely narrow line core (FWHM
$\sim$~6~km~s$^{-1}$) together with broad Lorentzian wings (FWHM
$\sim$~30~km~s$^{-1}$). The line core is consistent with only opacity
broadening of a thermal profile. The spatial distribution of CNM
linewidths is not random, but instead is extremely rich in
structure. High linewidths occur in distinct shell-like structures
with diameter of 100's of pc to kpc's, which show some correlation
with diffuse H$\alpha$ shells. The primary source of ``turbulent''
linewidth in the atomic ISM appears to be organized motions due to
localized energy injection on a scale of a few 100 pc.
\end{abstract}

\firstsection 
\section{A Brief Look at HI Thermodynamics}

Over the years a succession of eminent authors has considered the
question of which physical processes determine the temperature of
neutral atomic gas and which timescales are required to achieve (a
local) thermodynamic equilibrium. This began with Field, Goldsmith and
Habing (1969) and continued with Draine (1978), Shull and Woods (1985)
and most recently Wolfire {\it et al.} (1995). While some of the
processes involved are quite straightforward, others, like the
photoelectric emission from dust grains, have had to be substantially
updated to reflect our growing knowledge of the properties and
abundance of interstellar dust. The consensus which has emerged is
that heating is in fact dominated by the photoelectric heating from
small dust grains over a wide range of conditions and environments.
Cooling, on the other hand, is regulated primarily by emission in the
[CII] 158~$\mu$m fine structure line at densities in excess of about
1~cm$^{-3}$ and by Ly$\alpha$ emission at lower densities. The
relative importance of the various mechanisms is nicely illustrated in
Fig.~3 of Wolfire {\it et al.} (1995). In the same figure can be seen
the characteristic phase diagram for atomic gas. Two thermodynamically
stable phases are found. The first is the so-called Warm Neutral
Medium (WNM) which predominates at low densities and pressures, and
has a kinetic temperature that rises toward lower pressures from a
value of perhaps 5000~K to 10$^4$~K, with an accompanying increase in
ionization fraction, $x$~$\sim$~0.1 to 0.9. The second is the Cool
Neutral Medium (CNM) which is primarily found at high densities and
pressures, with a kinetic temperature that decreases towards higher
pressures from a value of some 200~K to perhaps as low as 20~K. Over some
range of intermediate pressures thought to be typical of the
interstellar medium in the local neighbourhood of the Galaxy
(P/k$\sim$ 2000~K~cm$^{-3}$), the two phases can coexist in pressure
equilibrium. However, given the strong gradient in the mid-plane
thermal pressure which is likely to follow from the combination of a
radial exponential stellar disk and a similarly declining gas surface
density distribution, we can safely predict that the inner regions of
disk galaxies will have predominantly rather cool CNM, while the outer
regions of galaxies will eventually be dominated by the WNM. At
intermediate radii we might also expect to see a radial increase in
the CNM kinetic temperature in response to the declining mid-plane
pressure. 

\section{The CNM -- WNM Phase Balance}

The phase balance of cool and warm neutral gas can be probed by
obtaining coupled observations of the HI emission and absorption along
the lines-of-sight to appropriate background sources. Observations of
the many bright background sources lying near the Galactic plane in
projection have provided a very good characterization of the atomic
gas properties in the solar neighbourhood. It turns out that there is
an excellent correlation between the observed absorption opacity and
the emission brightness derived from all of the existing high quality
data probing solar neighbourhood atomic gas. The distribution shown in
in Fig.~1a (from Braun and Walterbos 1992) is ``S-shaped'' and is
characterized by two asymptotic values. One is the maximum observed
brightness $T_B(\tau\rightarrow\infty)$~=~107~K, where the line
opacity becomes effectively infinite. The other, is the minimum
observed brightness $T_B(\tau\rightarrow0)$~=~4~K for which
vanishingly small opacities are still observed. The simplest physical
model which can account for this distribution, is one with a cool
component (the CNM) responsible for the majority of both emission and
absorption, which is embedded in a widely distributed non-absorbing
component (the WNM) that accounts for only a small fraction of the
emission linestrength, but does have a substantially wider linewidth.

\begin{figure} 
\centerline{
\epsfysize=6.3cm
\epsfbox[10 165 592 425]{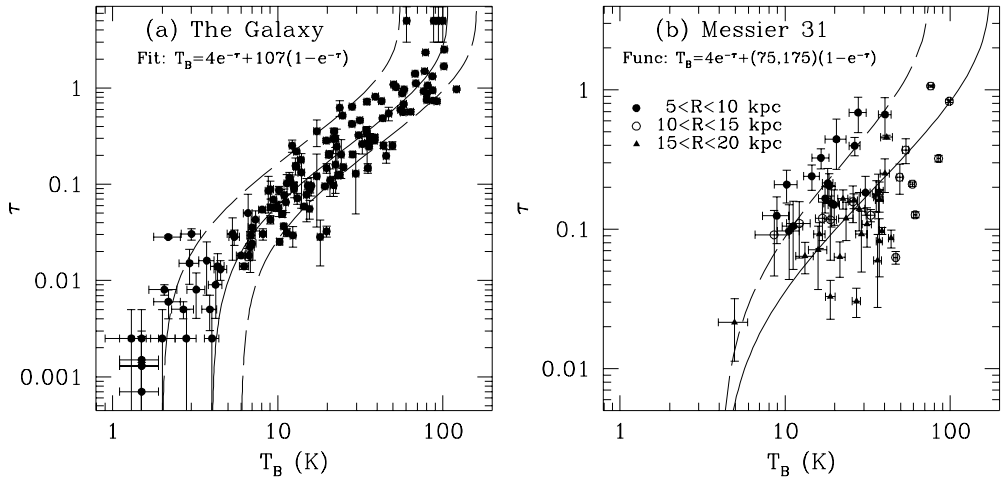}
}
  \caption{Parameterization of the HI opacity against emission
    brightness in the Galaxy and M31. (a) A two-component model fit to
    the Galactic data, together with an envelope at plus and minus
    50\% that contains most of the data points. (b) Illustration of a
    two-component model for M31, with characteristic temperatures of
    75 and 175~K for the inner and outer disks respectively. } 
\end{figure} 

Although a mean CNM kinetic temperature of about 110~K is indicated
for the solar neighbourhood of the Galaxy, it should be stressed that
the distribution has a scatter corresponding to about a factor of 2
local variation of the CNM temperature from the mean. This same degree
of temperature variation also seems to be indicated on smaller scales,
by the detailed differences between individual HI emission and absorption
profiles. Rather than being simple mirror images of eachother, as
would be the case for a truly isothermal component, the absorption
profiles are typically somewhat narrower than the corresponding
emission, suggesting somewhat cooler cores within a warmer mantle.
Solving for the actual column density ratio of the CNM to WNM in the solar
neighbourhood gives a value of about 3:1, which is substantially
higher than the value of about 1:1 which has been suggested in the past.

Comparably detailed analysis of the atomic phase balance in external
galaxies has proven much more difficult due to the very small number
of sufficiently bright background sources located by chance behind the
galactic disk. However, the external viewing perspective does provide
some advantages.  In the case of M31, it reveals a clear distinction
between the $\tau~-~T_B$ distributions observed at different
galactocentric radii through that system's disk (Braun and Walterbos
1992, see also Dickey and Brinks 1993).  Characteristic CNM kinetic
temperatures are observed to vary systematically from about 75~K at
5~kpc radius to 175~K at 15~kpc radius as shown in Fig.~1b. This is in
agreement with the expectation noted above to find a positive radial
gradient in the CNM kinetic temperature in response to a decrease in
the mean mid-plane pressure.

\section{The Two Observed Components of HI Emission}

A rather striking observational result of recent years is the
recognition that there are two observationally distinct components to
the HI emission seen from nearby galaxy disks (Braun 1995, 1997). When
sufficient physical and velocity resolution (of order 150~pc and
5~km~s$^{-1}$) are employed, the HI distribution is decomposed into:
(1) a distinct high brightness network (HBN) of emission features which
trace the regions of active star formation and (2) a diffuse
component of inter-arm and outer disk gas. 

\subsection{The High Brightness Network}

The observed HI emission brightness temperature of this component
varies between about 50 and 200~K. Since $T_B=T_S(1-e^{-\tau})$ and
the spin temperature, $T_S$, is equal to the kinetic temperature under
galactic conditions, these values of $T_B$ provide a lower limit to
the kinetic temperature of the gas. The visual appearance of this
component, in the case of relatively face-on galactic disks, is that
of a filamentary network, which is only marginally resolved in its
narrow dimension, but is extended over 100's of pc to kpc's in its
long dimension. In more edge-on systems, the network begins to fill
the entire disk as seen in projection. 

\begin{figure} 
\centerline{
\epsfysize=6cm
\epsfbox[1 175 560 615]{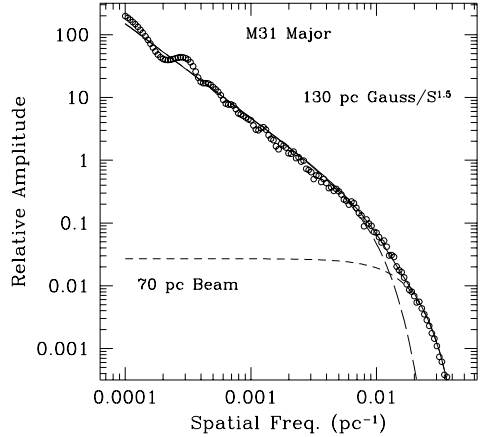}
}
  \caption{Amplitude spectrum of atomic structures near the North-East
    major axis of M31. The point spread function is a 70~pc FWHM
    Gaussian. The data is well-fit with a $-$1.5 index power law
    tapered with a 130~pc FWHM Gaussian in addition to the noise floor. } 
\end{figure} 

The best data on the physical extent of this component are those for
M31, where a physical resolution as high as 35~pc has been obtained,
with simultaneous good sampling of the spatial scales out to many kpc.
The amplitude spectrum for the NE major axis region of
M31 smoothed to 70~pc resolution is shown in Fig.~2 together with the
overlaid functional form of a power law with an index of $-$1.5
tapered by a 130~pc FWHM Gaussian and added to the measurement noise
floor.  Both the power law index as well as the characteristic
Gaussian FWHM are determined to about 10\% precision. The
corresponding power law index for a power-, rather than an amplitude
spectrum, would be $-$3.0.

The critical velocity resolution needed to at least marginally resolve
HI emission profiles appears to be about 5~km~s$^{-1}$. Velocity
smoothing from 5 to 10~km~s$^{-1}$ leads to a mean dilution of the
peak intensity of 15\% in M31 (Braun 1995), while smoothing from 1 to
5~km~s$^{-1}$ results in a mean peak dilution of only 3\% for Galactic
emission profiles with a peak brightness in excess of 80~K. 

\begin{figure} 
\centerline{
\epsfysize=5.6cm
\epsfbox[1 175 560 615]{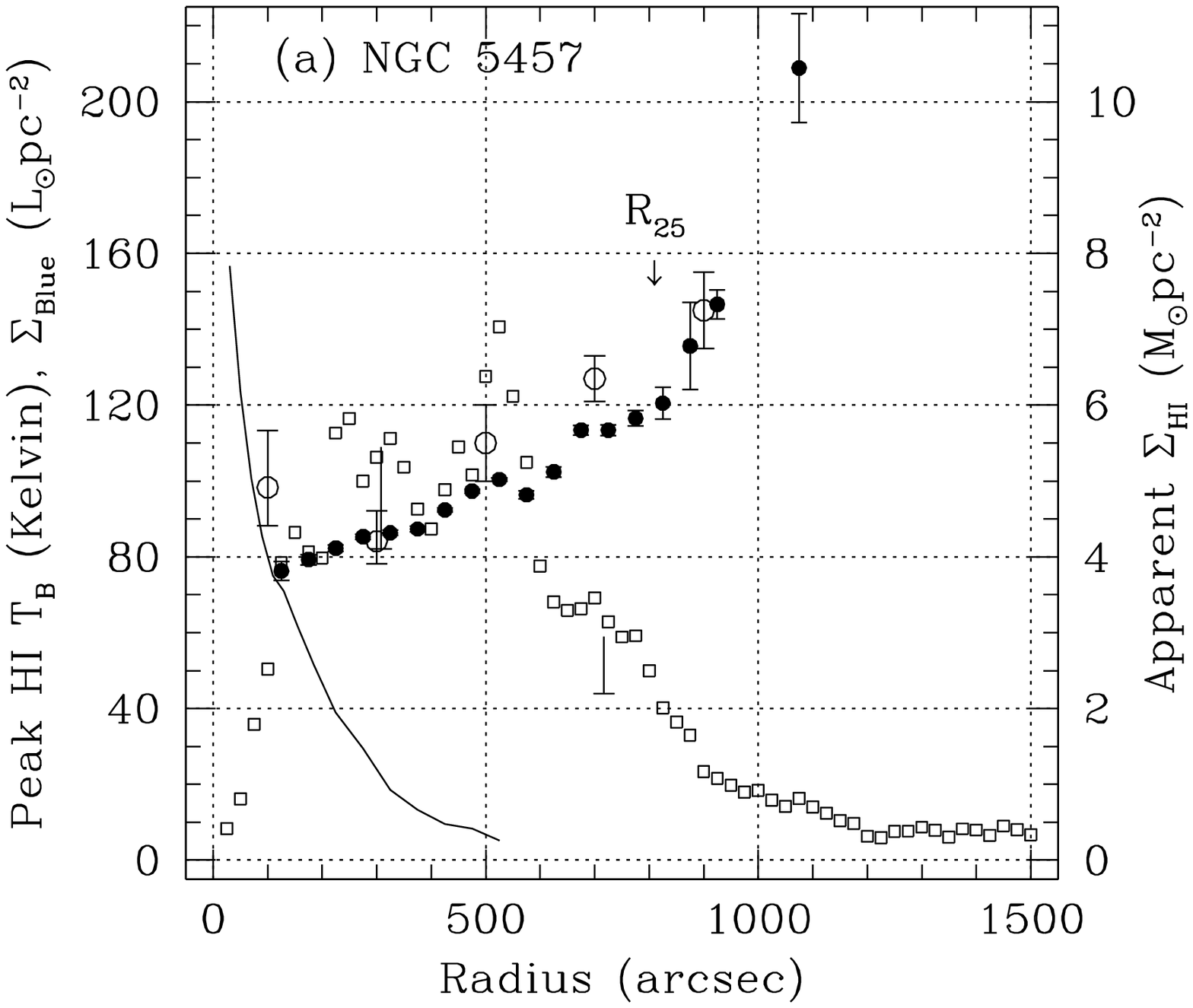}
\epsfysize=5.6cm
\epsfbox[40 175 600 615]{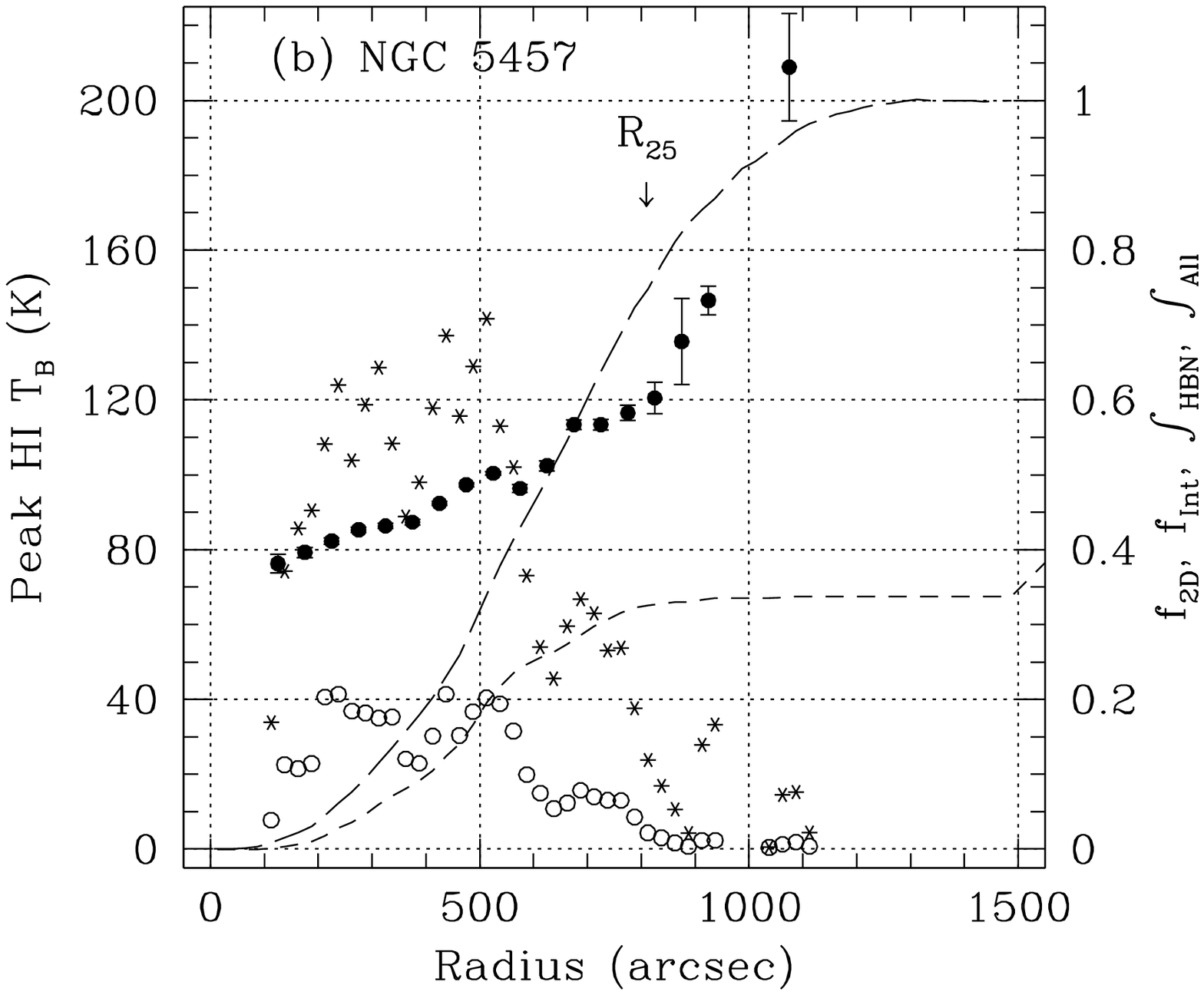}
}
  \caption{Radial profiles of CNM properties in
    NGC~5457. (a) Filled circles indicate peak brightness (80th
    percentile), open circles the kinetic temperature fit to the line
    profiles, squares the apparent face-on mass surface density, and
    the solid line is the face-on blue light surface brightness. Half
    error bars indicate the 3$\sigma$ lower limits to the mean spin
    temperature determined against background sources. (b) Filled
    circles indicate the peak brightness as in (a), open circles the
    face-on surface covering factor of the HBN, and the asterisks the
    local fraction of HI flux contained in the HBN. The short-dashed
    line is the accumulated flux in the HBN normalized to the total
    flux. The long-dashed line is the accumulated total flux.} 
\end{figure} 

The statistical properties of the HI HBN have been determined for
seven of the nearest spiral galaxies which have a moderate inclination
(Braun 1997). This component accounts for between 60 and 90\% of the
integrated HI emission inside of R$_{25}$ (the radius where the blue
optical light reaches a face-on surface brightness of
25~mag~arcsec$^{-2}$) and is concentrated to about 15\% of the disk
surface area.  Beyond R$_{25}$ this component abruptly disappears. It
seems clear that the HBN is a prerequisite for active star formation.
These distributions are illustrated in Fig.~3 for M101.

Another striking feature of the HBN, is the non-Gaussian form of the
line profiles. High signal-to-noise line profiles generated from
median co-aligned spectra within annular bins of galactocentric radius
have a very distinctive shape, as seen in Fig.~4a. An extremely narrow
line core (with FWHM less than about 6~km~s$^{-1}$) is superposed on
much broader line wings with a profile that is well-fit with a
Lorentzian with a FWHM of about 30~km~s$^{-1}$. Similar line profiles
have been detected in nearby dwarf galaxies by Young and Lo (1997).
The narrow line cores provide a strong upper limit to the HBN kinetic
temperature of 300~K, while the observed emission brightness provides
a direct lower limit of 50 to 200~K. The HBN kinetic temperature is
thus very well-constrained, and it allows unambiguous identification
of this component with the Cool Neutral Medium of the ISM. A close
correspondence of the CNM zone with star formation activity was
suggested by Elmegreen and Parravano (1994).

\begin{figure} 
\centerline{
\epsfysize=6.3cm
\epsfbox[50 200 540 625]{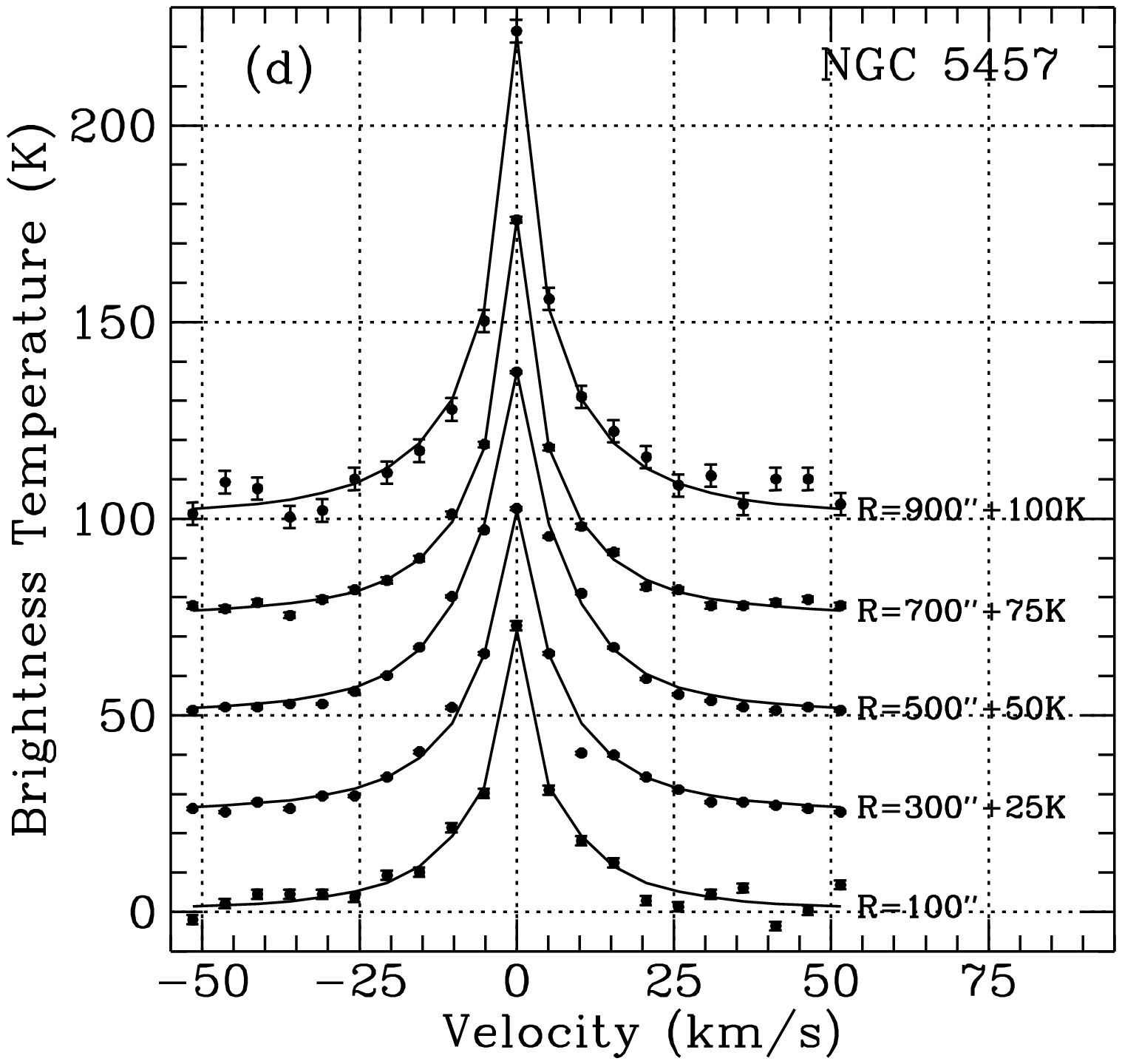}
\epsfysize=6.3cm
\epsfbox[56 113 576 678]{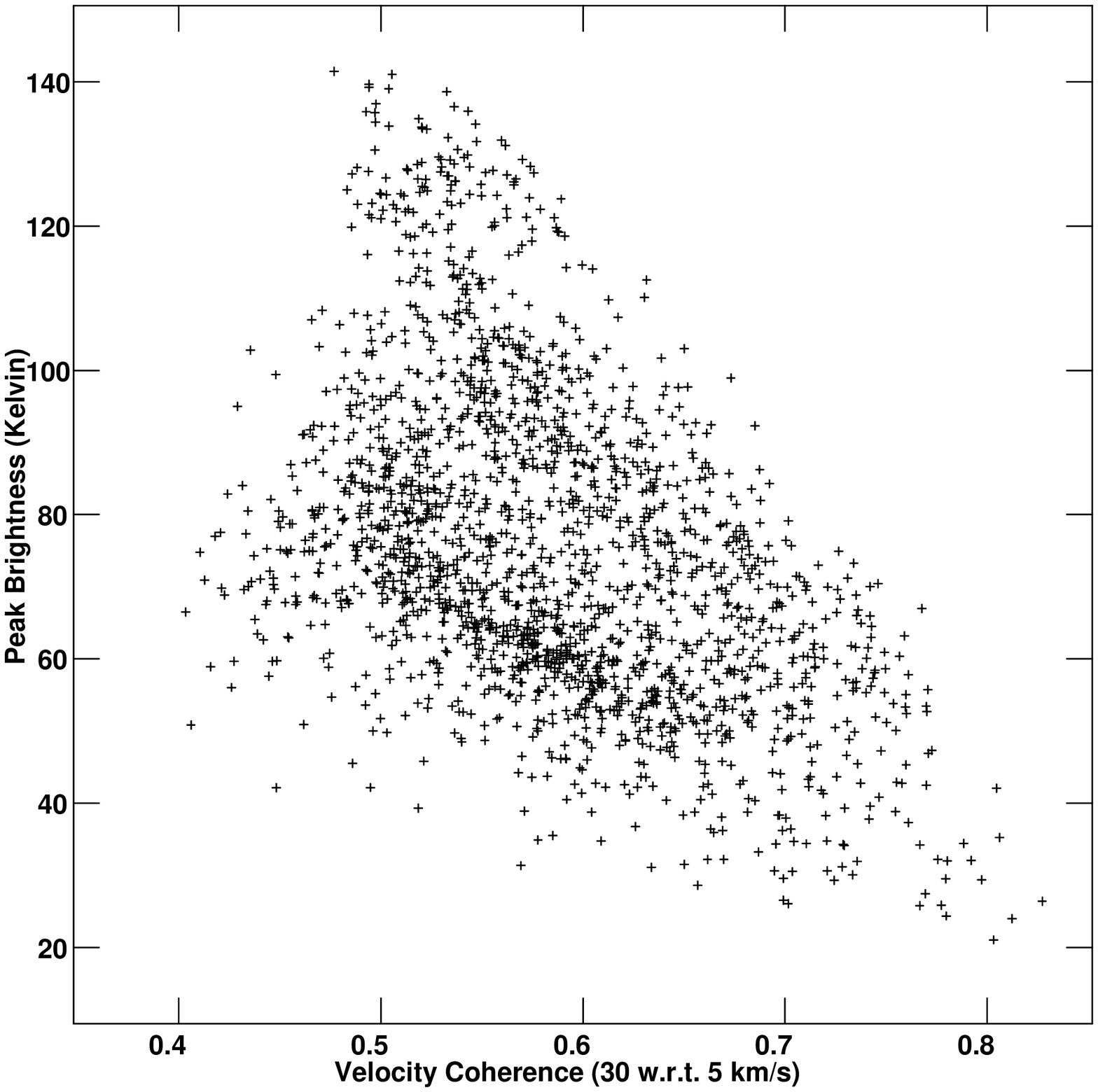}
}
  \caption{Line profile properties of the CNM. (a) Median co-aligned
    emission line profiles in NGC~5457 of HI brightness temperature
    within annular elliptical zones at the indicated radii. Solid
    lines are a two component radiative transfer fit. (b) Scatter plot
    of peak brightness of a profile against VC$_{30km/s}$ (a measure
    of the linewidth explained in the text), for the North-East major
    axis of M31.}
\end{figure} 

The entire sample of observed nearby galaxies shows the systematic
trend for an increasing peak brightness temperature of the CNM with
radius, as illustrated in Fig.~5a. Furthermore, the different observed
galaxy types form a nested system when plotting peak observed
brightness temperature against radius, in the sense that later type
spirals are offset to higher peak brightness temperatures at all radii
than earlier type spirals.

\begin{figure} 
\centerline{
\epsfysize=6.3cm
\epsfbox[10 165 592 425]{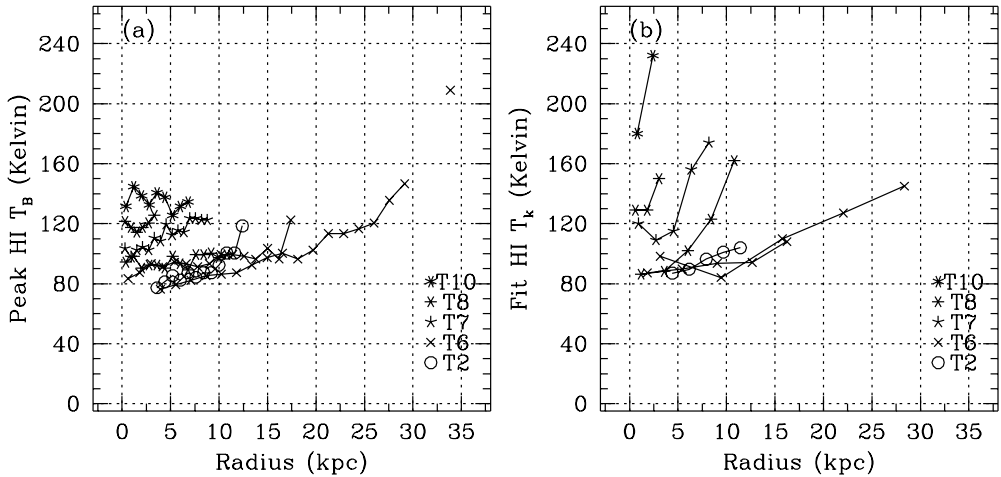}
}
  \caption{Radial profiles of CNM peak brightness in (a) and derived kinetic
    temperature in (b) for seven nearby galaxies. Different symbols
    are used to indicate the morphological type of each galaxy.}
\end{figure}

Fitting of the median line profiles with a simple two component
radiative transfer model allows derivation of gas kinetic temperatures
and column densities as function of galactocentric radius. Kinetic
temperatures closely track the observed brightness temperatures within
inner galaxy disks where HI opacties are typically quite high. The
kinetic temperature systematically increases to larger radii as shown
in Fig.~5b, with a corresponding decline in the HI opacity. Just as we
discussed in the preceeding section, the theoretical expectation seems
to be borne out, that there is a clear gradient in the CNM kinetic
temperature with radius which is very likely tracking the decreasing
mid-plane thermal pressure profile of galactic disks.

\subsection{The Low Brightness Disk}

As outlined above, most of the HI emission within R$_{25}$ is due to
the high brightness filamentary network of the CNM. However, it is
still the case that half of the integrated HI emission
arises within a diffuse disk component. This is illustrated in Fig.~3b,
where the accumulated HI flux is plotted as function of radius for both
the HBN (i.e. CNM) component and the total emission. While the HBN
abruptly disappears near R$_{25}$, the total flux still rises by about
a factor of two. On average, diffuse HI emission accounts for only
20\% of the HI flux, but represents some 85\% of the face-on surface
area within R$_{25}$, while outside of R$_{25}$ it dominates.

Since the diffuse component has an extremely low brightness
temperature in emission, it is very difficult to determine it's
properties in detail. Given the typical measurement sensitivity, it is
necessary to adopt spatial smoothing to kpc scales to allow direct
detection of the emission. With a spatial resolution as coarse as a
kpc, kinematic effects begin to have a major influence on the shape of
emission line profiles. Such beam smearing substantially complicates
the physical interpretation. The cleanest data currently available is
that for the diffuse disk of M101 (Fig.~10 of Braun 1997). These line
profiles are consistent with a dominant component of WNM with little
additional broadening over the thermal linewidth (24~km~s$^{-1}$ FWHM)
of a 10$^4$~K gas.  This consistency does not provide a proof of such
an identification, but it does make it plausible.

\section{The Origin of HI Linewidth}

\begin{figure} 
\centerline{
\epsfysize=17.5cm
\epsfbox[133 148 478 645]{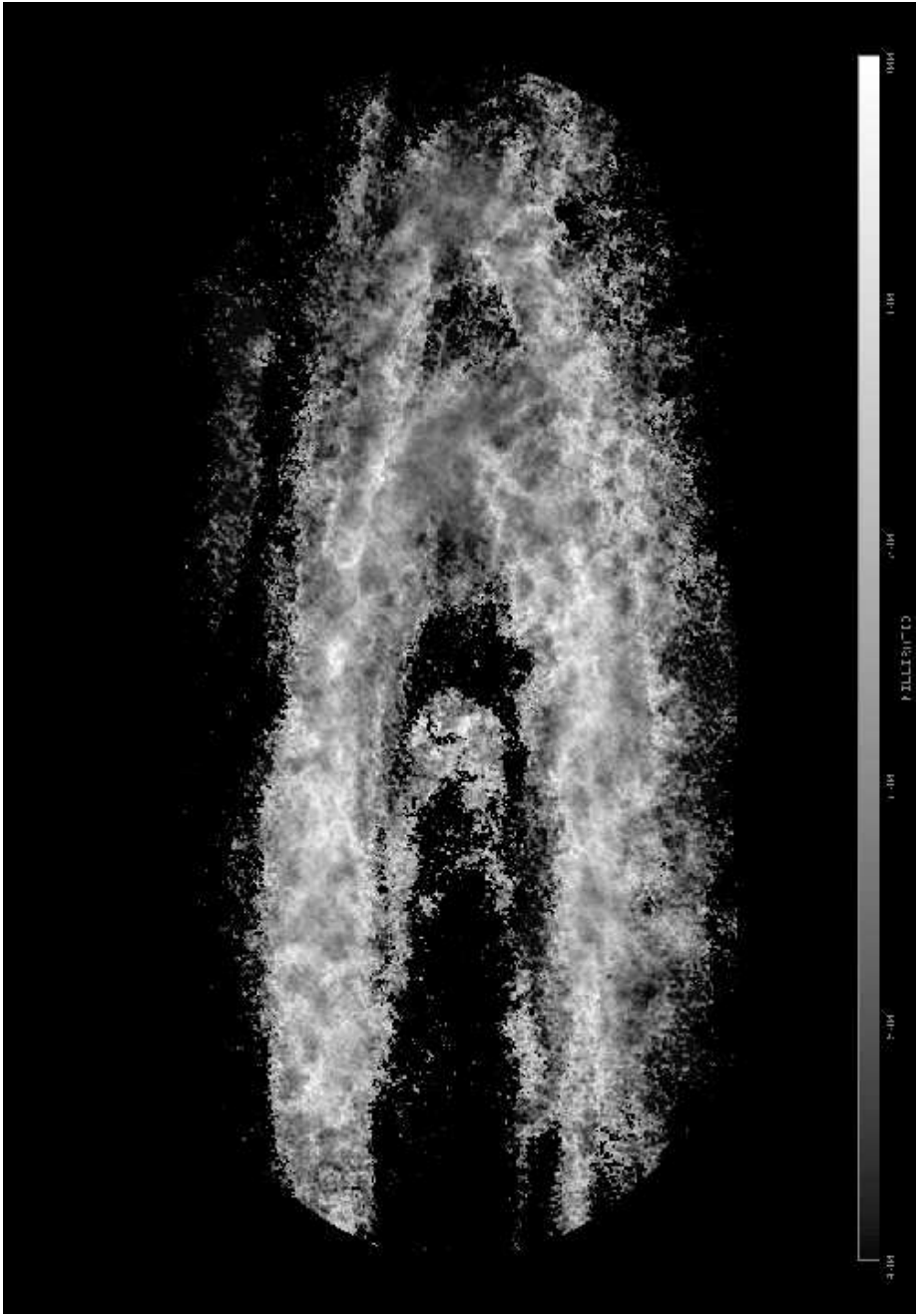}
}
  \caption{The distribution of HI linewidth in the North-East half of
    M31. The linewidth measure, VC$_{30km/s}$ (defined in the text)
    is observed to vary between about 0.4 and 1.0}
\end{figure}

Having provided some background relevant to the physical state of the
atomic gas in spiral galaxies, we can turn to an issue which is
perhaps more directly related to the topic of this volume, namely,
what is the primary mechanism which is responsible for the observed
linewidth?  In the past, moderate spatial and velocity resolution, of
perhaps 1~kpc and 10~km~s$^{-1}$, have typically been employed for HI
imaging studies of nearby galaxies. Line profiles were usually fit
with a single Gaussian component and the resulting linewidths were
averaged over large regions to derive estimates of the velocity
dispersion, $\sigma~\sim$~8~km~s$^{-1}$ ($\sigma$~=~FWHM/2.36), with
some indication for an increase at large radii. By obtaining much
higher resolution data, especially in a nearby system, it is possible
to address the issue of linewidth in considerably more detail. In
Fig.~6 we present an image of the HI linewidth in the North-East half
of M31. Rather than fitting any assumed functional form to individual
spectra, an image was made of the ``velocity coherence'', defined as
VC$_{30km/s}(x,y)~=~P_{30km/s}(x,y)/P(_{5km/s}x,y)$, where
$P_{Rkm/s}(x,y)$ is the peak brightness observed at a spectral
resolution of $R$~km~s$^{-1}$. Those lines-of-sight with profiles
which are 30 or more km~s$^{-1}$ in width have a value of unity in
this image, while the narrowest profiles result in values of
VC$_{30km/s}$~$\sim$~0.4. What is striking in this image is the wealth of
structure that is apparent. 

High velocity width is not a property which is randomly distributed
over a galactic disk. Instead, regions of high velocity width occur in
organized structures with a scale-size of some 100's of pc to kpc's
and which are reminiscent of filamentary cavity walls. Such high
linewidth ``shells'' are quite distinct from quiescent pockets, with
similar dimensions, in which the line core is extremely narrow. As we
showed previously, (in Fig.~4a) the median line profiles of the CNM can
be well modeled with a combination of a purely thermal component (of
80 to 150~K) with only modest broadening by opacity effects, together
with broad Lorentzian line wings. Now it becomes clear that even the
core and wing components are to some extent structurally distinct.

Further insight into the processes at work comes from considering the
relationship between peak line strength and the linewidth. In Fig.~4b
we plot these values for a 2.5~kpc square region near the M31 NE major
axis crossing at 10~kpc radius. Peak brightnesses at this radius vary
by roughly a factor of two, between 80 and 140~K, and occur at the
minimum velocity coherence of about VC$_{30km/s}$~=~0.5. This
underlines the point made earlier, that at a given galactocentric
radius, intrinsic kinetic temperature variations of about a factor of
two are present.  As the linewidth increases, the peak brightness
declines systematically. The distribution has a wedge-like appearance,
in which the factor of two (brightness) temperature variation is
retained at all linewidths. This is suggestive of a direct
correspondence between the linewidth and the HI opacity. Since we
expect a typical opacity at this radius in M31 of about unity (much
like the Solar Neighbourhood of the Galaxy), a linearly increasing
linewidth should yield a linearly decreasing peak emission brightness
temperature for a constant gas column density.

Some idea of a possible causal agent for the high linewidth structures
comes from the comparison of the image in Fig.~6 with one of H$\alpha$
emission. While there is certainly no one-to-one correspondence, there
are several notable instances in which diffuse, large diameter
H$\alpha$ super-shells have high linewidth HI counterparts.  In view
of the rather short expected lifetimes for H$\alpha$ emission due to
localized massive star formation (less than about 5~Myr) compared to
the timescale for super-shell dissipation in the ISM (more than about
30~Myr) it seems plausible that what we are seeing is the fossil
signatures of localized energy injection integrated over the past
50~Myr or so.

Independent of the particular causal agent, it is quite clear that
organized motions on 100 to 1000~pc scales play a dominant role in
determining the line profile of atomic gas. These motions, together
with the purely thermal linewidths of the CNM seem to provide a good
description of the star-forming disks of spiral galaxies. 
 
\begin{acknowledgments}

\end{acknowledgments}

\end{document}